\documentclass{optica-article}

\journal{ome}

\articletype{Research Article}
\usepackage{siunitx}
\usepackage{lineno}
\usepackage{amsmath}

\begin{document}

\title{Single-point-diamond-turned GaAs disk resonator with over a million optical quality factor}

\author{Lily M.\ Platt,\authormark{1,2} Mallika Irene Suresh,\authormark{1,2,*} Farhan Azeem,\authormark{1,2} Luke S.\ Trainor,\authormark{1,2} and Harald G.\ L.\ Schwefel\authormark{1,2}}

\address{\authormark{1}The Dodd-Walls Centre for Photonic and Quantum Technologies, New Zealand\\
\authormark{2}Department of Physics, University of Otago, Dunedin 9016, New Zealand}

\email{\authormark{*}mallika.suresh@otago.ac.nz} 



\begin{abstract}
Gallium arsenide optical resonators have been identified as platforms for light-matter interactions ranging from optomechanics to quantum electrodynamics involving nonlinear optics. Here, we present a 5-mm-diameter whispering gallery mode resonator made of undoped gallium arsenide. The fabrication was done using single-point diamond turning followed by polishing with diamond solutions of varying grain size. The resulting resonator was optically characterized to have a quality factor of more than a million.
\end{abstract}

\section{Introduction}
Gallium arsenide (GaAs) has a number of properties that have made it an interesting material in the field of optics; its broad transparency window in the infrared from \SI{0.9}{\um} to \SI{17}{\um} has made it useful for applications such as spectroscopy~\cite{haas_gallium_2020}, sensing~\cite{chen_on-chip_2020} and imaging~\cite{geum_monolithic_2019}. It has been used to implement optical switches~\cite{van_all-optical_2002, ravindran_gaas_2012} and filters~\cite{chin_gaas_1999}, which also benefit from its direct bandgap and high electron mobility, two properties that have historically made it useful for solar cells and photovoltaics~\cite{bosi_potential_2007, bett_iii-v_1999}. The higher electron mobility and direct bandgap in comparison to silicon (Si) make GaAs a good platform for optoelectronics~\cite{wada_optoelectronic_1988}. Due to their broad transparency and large nonlinear optical susceptibilities, GaAs, and other similar zincblende semiconductor crystals like GaP and ZnSe, are important candidates for nonlinear optical mixing process for frequency generation~\cite{wang_quasi-phase_2019, otman_phase_2018, sotor_all-fiber_2018}, optical parametric oscillation~\cite{fu_high-beam-quality_2019}, optical parametric generation~\cite{becheker_optical_2022}, and lasing~\cite{alanis_optical_2019, zhang_ultralow_2021}. 

The isotropic linear optical properties due to the cubic symmetry have made it necessary to implement quasi-phase matching by orientation patterning to achieve reasonable efficiency of nonlinear optical processes in GaAs~\cite{grisard_quasi-phase-matched_2012,gonzalez_second-harmonic_2013,tanimoto_quasi-phase-matching_2021}. However, the structure of such $\bar{4}$ symmetric crystals entails an inversion of the crystallographic domains with a \ang{90} rotation around a $\bar{4}$-axis. Therefore, as discussed and demonstrated in~\cite{kuo_4-quasi-phase-matched_2009, kuo_second-harmonic_2014}, shaping GaAs into whispering gallery mode resonators (WGMRs) with the axis of rotation along a $\bar{4}$-axis provides an inherent quasi-phase matching mechanism without engineered domain inversions such as by orientation patterning. Beyond the typical advantage that resonant cavities offer by enhancing nonlinear optical interaction between light and matter which benefits frequency mixing~\cite{andronico_difference_2008, chang_strong_2019} and electro-optomechanical effects~\cite{ allain_electro-optomechanical_2021}, GaAs resonators have been used to demonstrate efficient frequency generation by inherent quasi-phase matching~\cite{kuo_on-and-off_2011, kuo_second-harmonic_2014, kuo_mixing_2018}. 

Due to its extensive use in the electronics industry as well as the significant interest in developing integrated photonic chip devices, established methods of fabrication such as electron-beam lithography and wet-etching~\cite{chin_gaas_1999, baker_critical_2011, chen_on-chip_2020}, metal organic chemical vapour deposition~\cite{yoon_gaas_2010}, molecular beam epitaxy~\cite{wirths_room-temperature_2018} etc., have been borrowed for use in developing photonic platforms and chip-based optoelectronic integrated circuits. For instance, an on-chip disk-shaped GaAs WGMR with self-assembled quantum dots has been used for demonstrating coherent quantum rerouting of photons~\cite{brooks_integrated_2021}. However, it is difficult to achieve very smooth surfaces with such techniques and the quality factors ($Q$) of the fabricated GaAs resonators are lower than the expected intrinsic $Q$-factor based on the absorption losses in the material. To the best of our knowledge, Chang et al., have obtained the highest $Q$-factor in the field, using an integrated GaAs ring-shaped resonator (radius = \SI{100}{\um}), which attained an intrinsic $Q$-factor of \num{2.6e5} at \SI{2}{\um} and \num{5.7e4} at \SI{1}{\um}~\cite{chang_strong_2019}. Guha et al., reported an intrinsic $Q$-factor value of \num{6e6} with a sub-wavelength-thick GaAs resonator coated with a fine layer of alumina to decrease the surface scattering losses~\cite{guha_surface-enhanced_2017}. Without the atomic layer deposition of alumina, their $Q$-factors are an order of magnitude smaller.

Here, we present a disk-shaped WGMR with a radius of \SI{2.3}{\milli\meter} fabricated using single-point diamond turning followed by hand-polishing with diamond slurry. To the best of our knowledge, this is one of the first optical measurements performed on a GaAs WGMR fabricated with the mechanical technique of single-point diamond turning. The loaded $Q$-factor of the fabricated WGMR at telecom wavelengths (\SI{1550}{\nano \meter}) was measured to be \num{1.1e6}. 

\section{Fabrication}
The fabrication of the WGMR investigated in this paper was performed using a rotational lathe and a diamond tool using a method called single-point diamond turning~\cite{floriansthesis}. Thermal methods of fabrication are avoided to keep the crystal structure intact and therefore, ensure that the nonlinearity of the crystal is preserved. Furthermore, single-point diamond turning has been found to be a favourable method for generating nano-smooth surfaces with good control over shape definition and accuracy when compared with electrochemical wetstamping and laser machining, while also saving on fabrication time in contrast to methods such as lapping and chemical polishing~\cite{chen_fundamental_2020}. This method has also been successfully used on brittle materials like GaAs, Si, silicon carbide, etc.

\begin{figure}[h!]
    \centering
    \includegraphics[scale=0.6]{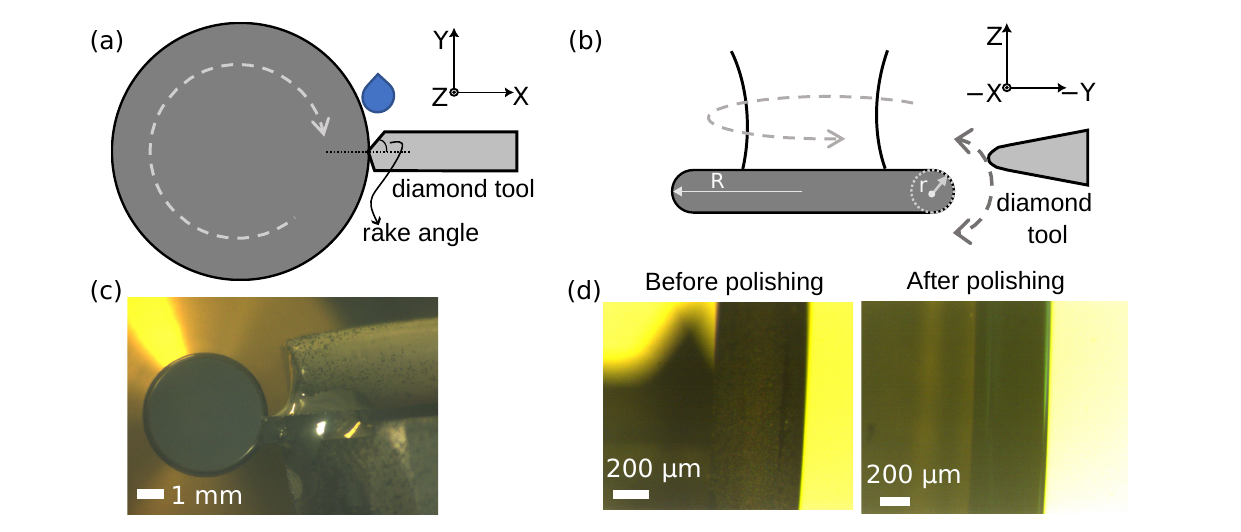}
    \caption{Fabrication of a GaAs WGMR using single-point diamond turning with water dripped on the cutting position: (a) and (b) depict schematics of the WGMR spinning around its axis in the X-Y plane and the Y-Z plane respectively; (c) and (d) are photographs taken during the cutting and polishing respectively.}
    \label{fig:fabrication}
\end{figure}

A circular blank was initially drilled out of a GaAs wafer (from UniversityWafer, Inc., semi-insulating with a resistivity of $>\SI{e7}{\ohm \centi\meter}$) using a tube-shaped brass drill bit spinning around its rotational axis on a lathe. \SI{30}{\um} diamond slurry (an oil- or water- based suspension of diamond particles of a specific size) was continually dropped between the wafer and drill bit. Single-point diamond turning was then used to cut the cylindrical disk down to a well-defined curved edge. This process consisted of placing the disk on the lathe which was set spinning at \SI{100}{\hertz}, while a diamond tool (whose tip has a radius of curvature of \SI{0.18}{\milli \meter}) is brought into single-point contact with the rim of disk. 

To ensure a fine surface quality during the diamond cutting process, the cut depth of each step during cutting and rake angle should be considered. The rake angle is the optimal angle at which the diamond should be placed relative to the spinning piece of wafer to cause less strain on the wafer surface. This angle is illustrated in Fig.~\ref{fig:fabrication}(a). As seen in Fig.~\ref{fig:fabrication}(b), the diamond tool moves on a predetermined path to carve out the desired geometry of the WGMR. The cutting parameters used were based on investigations reported in~\cite{chen_fundamental_2020} for ductile machining of a GaAs wafer. Our rake angle of \ang{-50} was chosen to be close to the \ang{-40} used there, but still allowed ample clearance with the diamond tool's built-in rake angle of \ang{-45}. The feed rate was \SI{1}{\um/rev}. The cutting depth was limited by the translation stages used and was not optimized. However, it has been noted that ductile machining of (001) GaAs was challenging even at very low cut depths~\cite{chen_fundamental_2020}. The major radius $R$ and the minor radius $r$ (shown in Fig.~\ref{fig:fabrication}(b)) were designed to be \SI{2.54}{\milli \meter} and \SI{0.112}{\milli \meter} respectively.

As the diamond tool cuts the GaAs wafer, caution must be taken to ensure no accidental inhalation of the GaAs dust produced by the cutting process. According to the National Institute for Occupational Safety and Health (NIOSH), GaAs should be regarded as a potential carcinogen. As a precaution during the cutting process, water is dripped on the position where the wafer is being cut as seen in the photograph in Fig.~\ref{fig:fabrication}(c), and the setup is enclosed in a plastic box. The water dripping allows safe collection of the GaAs dust in a container below the lathe, which is later disposed of appropriately.
Additionally, the water acts as a cutting lubricant.

Polishing of the rim of the WGMR is usually required to obtain a smoother surface quality after cutting, which is especially important because of the low fracture toughness and strong anisotropy of GaAs~\cite{chen_fundamental_2020}. Polishing of optical crystals is commonly performed by hand in order to have control over the stress exerted on the sample and to target particular parts that might need more polishing than others. Therefore, following the cutting, the curved edge of the WGMR is polished using an optical tissue with diamond slurry solutions varying from \SI{9}{\um} to \SI{1}{\um}. To ensure that an even distribution of slurry was applied to the spinning WGMR, the tissue was moved randomly during the polishing process, and the direction of rotation was reversed periodically. When the surface that is being polished only has scratches from the diamond slurry size, a diamond slurry of smaller size should be used, after thoroughly cleaning the polished surface to remove any remnants of the larger diamond pieces. The polishing process is thus repeated until the WGMR surface becomes very reflective when brightly illuminated and minimal scratches could be observed through a microscope, as can be seen in Fig.~\ref{fig:fabrication}(d). As diamond slurry was used in consecutively descending sizes, it was crucial to properly clean the WGMR with isopropanol on an optical tissue between each polishing step with different slurry sizes to avoid cross contamination.

\section{Optical characterisation}
\begin{figure}[tb]
    \centering
    \includegraphics[scale=0.48]{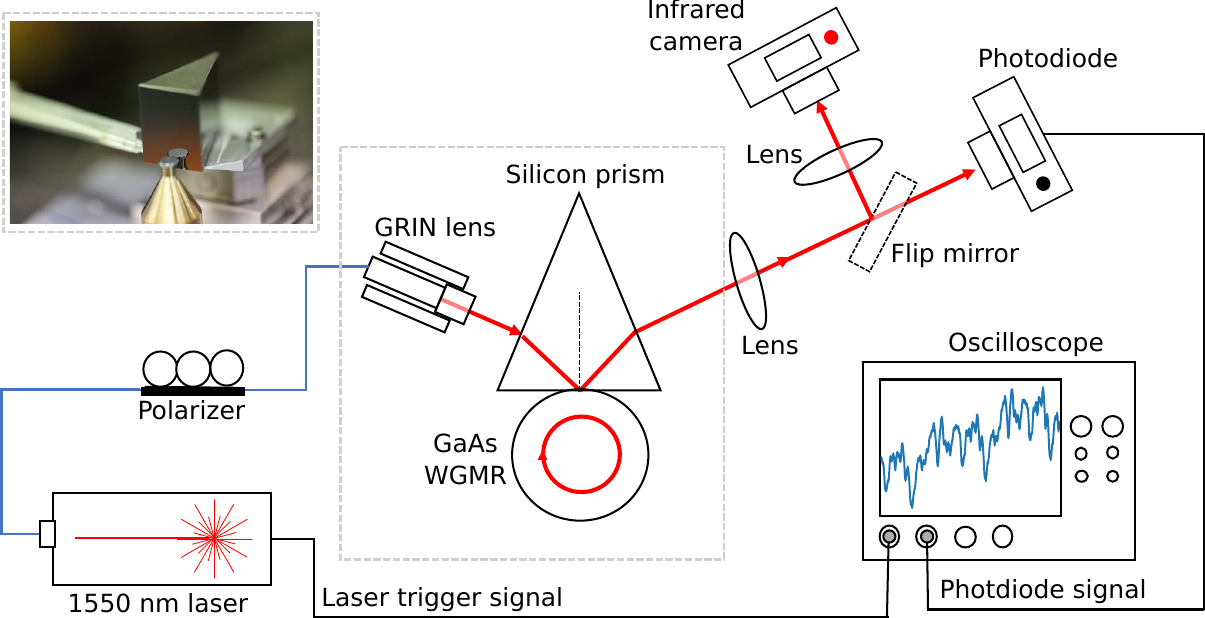}
    \caption{Schematic of the experimental setup to determine the $Q$-factor of the GaAs WGMR: \SI{1550}{\nano \meter} laser beam is sent via an optical fibre to a 3-paddle-polarization controller and then focused using a graded index (GRIN) lens into the prism. The light is reflected off the base of the prism. Through evanescent excitation, light is coupled into and out of the WGMR. The emerging light is collected and focused by a lens to an infrared camera or photodiode to be imaged or measured respectively. Inset: Photograph of the Si prism and GaAs WGMR.}
    \label{fig:setup_diagram}
\end{figure}

Optical characterisation tests of the GaAs WGMR were performed at telecom wavelengths ($\lambda$=\SI{1550}{\nano \meter}) using the setup in Fig.~\ref{fig:setup_diagram}. Such measurements can give an estimation of whether the quality of the fabrication is limited by the surface roughness (in which case further polishing might be required) or the absorption of the light by the bulk material itself.

To couple light into and out of the WGMR, the prism-coupling technique was used ~\cite{azeem_active_2022,couplingviaprism,near_fieldcoupling}. The WGMR is placed close to the middle of the base of an isosceles prism. A laser beam is then directed towards one side of the prism, the beam refracts at the prism face, gets reflected at the centre of the inside base of the prism (designed to be at the point closest to the WGMR), and then exits symmetrically through the opposite side of the isosceles prism. The prism material is chosen to be silicon (Si) since its refractive index, $n_{\text{Si}} = 3.48$ ~\cite{mccaulley_temperature_1994}, is larger than that of the WGMR $n_{\text{GaAs}} = 3.37$~\cite{mccaulley_temperature_1994} at the excitation wavelength. Therefore, coupling to the modes can occur via frustrated total internal reflection.

\begin{figure}[h!]
    \centering
    \includegraphics[scale=0.6]{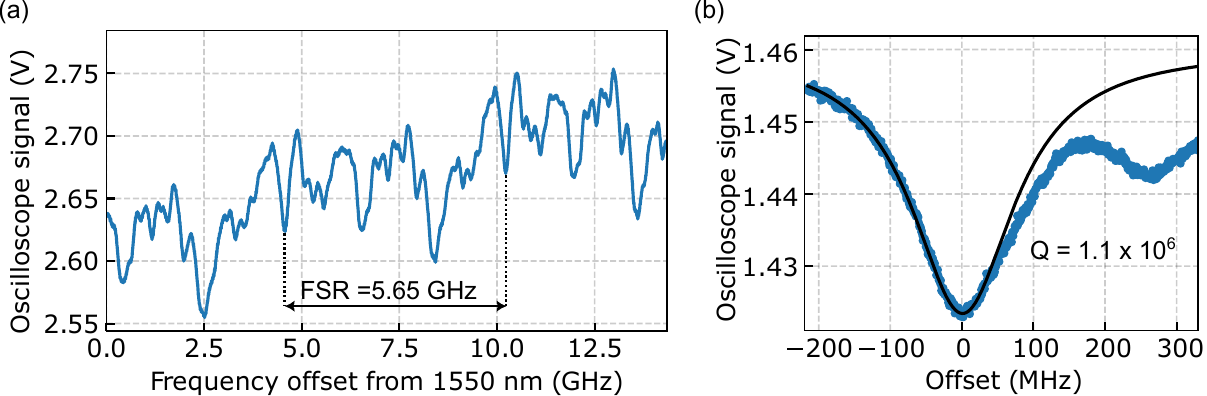}
    \caption{Spectra showing the WGMs excited in the GaAs disk WGMR: (a) shows part of the transmission spectrum observed on the oscilloscope with the FSR labelled. (b) shows a plot of a single mode  and the black line is the Lorentzian fit of the data, from which the linewidth of \SI{170}{\mega\hertz} is obtained. The $Q$-factor of this mode is calculated to be $1.1 \times 10^{6}$. }
    \label{fig:first_modes}
\end{figure}

Next, dips in the transmission spectrum were observed and were identified as WGMs. A high-resolution optical spectrum analyzer (Finisar WaveAnalyzer 1500s) was used to measure the frequency range of the laser's piezo sweep. The spectral spacing between modes of the same mode family gives the free spectral range (FSR) of the WGMR. This parameter is given by $\text{FSR} = c/(2\pi n_\text{GaAs}R)$ for the lowest-order radial mode, which using $R$ = \SI{2.54}{mm} and $n_\text{GaAs}$ = 3.37, gives $\text{FSR} = \SI{5.57}{\giga\hertz}$. As can be seen from Fig.~\ref{fig:first_modes}(a), the measured FSR is \SI{5.65}{\giga\hertz}. The difference can be attributed to an inaccurate frequency calibration, a possible difference in the refractive index of the GaAs wafer used or that the mode being evaluated is a higher-order radial mode. 

For the $Q$-factor measurement, we needed a more accurate mapping of the horizontal axis of the oscilloscope to the frequency sweep of the laser over a GHz. The laser signal is modulated using a phase modulator with a fixed frequency (\SI{100}{\mega\hertz}). This results in sidebands at \SI{100}{\mega\hertz} on either side of the signal, allowing a mapping of the horizontal axis based on this known frequency separation~\cite{azeem_active_2022}. The $Q$-factor was obtained by fitting a Lorentzian function to a particular mode and extracting the linewidth from the fit. The linewidth was found to be $\Delta f$ = \SI{170}{\mega \hertz}, from which the $Q$-factor at these wavelengths (i.e., centre frequency of the mode $f_0 \approx$ \SI{193}{\tera\hertz}) can be calculated as $Q$ = $f_0/\Delta f$ = \SI{1.1e6}{}. This value corresponds to a loaded $Q$-factor as the effect of the coupling prism is not negligible. This indicates that the intrinsic $Q$-factor of the WGMR would be higher than 1.1 million. To the best of our knowledge, this is more than five times higher than existing state-of-the-art intrinsic $Q$-factor in a GaAs WGMR~\cite{chang_strong_2019}. 

The optical $Q$-factor is dependent on the following three loss mechanisms: surface scattering losses $Q_{\text{ss}}$, material absorption losses $Q_{\text{abs}}$ and radiative losses $Q_{\text{rad}}$ in such WGMRs. The theoretical $Q$-factor is therefore given by: $Q^{-1} = Q_{\text{ss}}^{-1} + Q_{\text{abs}}^{-1} + Q_{\text{rad}}^{-1}$. In such mm-sized WGMRs with good circular symmetry, the $1/Q_{\text{rad}}$ is negligible. Surface scattering affects the $Q$ as $Q_{\text{ss}}\approx3\lambda^3R/(8\pi^2 n_{\text{GaAs}}B^2\sigma^2)$~\cite{gorodetsky_rayleigh_2000}, where $\lambda=\SI{1550}{\nano \meter}$ and $\sigma$ and $B$ are the surface roughness and the correlation length of the roughness. With $\sigma=B = \SI{80}{\nano\meter}$, we get $Q_{\text{ss}}\sim \SI{2.6e6}{}$. The bulk absorption limited $Q$ is given by $Q_{\text{abs}}\approx 2\pi n_{\text{GaAs}}/(\lambda\alpha)$, where $\alpha=\SI{0.065}{cm^{-1}}$~\cite{khan_optical_2007-1} is the intensity attenuation coefficient. This gives $Q_{\text{abs}} = \SI{2.1e6}{}$. Calculating the overall $Q$ from these numbers, we get $\SI{1.1e6}{}$, which is the measured $Q$ of the fabricated WGMR. This calculation indicates that our $Q$ measurement is close to absorption limited; the additional loss could be due to such a level of surface roughness or possibly additional absorption from a different purity of GaAs crystal. 
We were unable to verify that the experimentally measured $Q$ was taken at critical coupling due to the crowded spectrum; if this measurement was at critical coupling, the intrinsic $Q$ of the WGMR would be twice that of the loaded $Q$, i.e., \SI{2.2e6}{}. This would indicate that the fabricated WGMR was absorption-limited and the scattering losses due to the surface roughness were negligible. 

\section{Conclusion}
Here, we presented a disk-shaped GaAs WGMR fabricated via single-point diamond turning. Polishing with \SI{1}{\um} diamond slurry results in an optical quality factor of the resonator which is over one million at \SI{1550}{\nano \meter}. To the best of our knowledge, this $Q$-factor is highter than the previously reported value in a GaAs WGMR~\cite{chang_strong_2019}. As efficiency of any process occurring in such a resonator would scale with the $Q$-factor, the advantage brought by such an improvement of optical quality in GaAs WGMR is evident. Chip-based GaAs microresonators have been used in photonic signal processing as all-optical switches~\cite{lin_all-optical_2014} and optomechanical frequency combs~\cite{allain_electro-optomechanical_2021}. Furthermore, such semiconductor-based WGMR can be fabricated as monolithically integrated modules on silicon platforms for room-temperature lasing~\cite{wirths_room-temperature_2018} or high-sensitivity temperature sensors~\cite{chen_on-chip_2020}, where material properties like the direct band gap and large thermo-optic coefficient, are further enhanced by the resonator geometry. Similarly, low-threshold nonlinear optics could be demonstrated by exploiting the resonator geometry to amplify the nonlinear optical properties of GaAs~\cite{chang_strong_2019} and, the inherent quasi-phase matching resulting from the geometry of a GaAs-based WGMR is an important advantage for nonlinear optical processes as seen in~\cite{roland_frequency_2020, kuo_second-harmonic_2014}. Furthermore, GaAs has been identified as an interesting material for quantum photonics as well, as described in~\cite{boikov_nonlinear_2024}, where the the higher free electron recombination rate in comparison to silicon makes it perform better for quantum computation. 

\begin{backmatter}
\bmsection{Funding} This project is funded by an MBIE Endeavour Fund - Smart Ideas (UOOX2106) and MBIE Catalyst:Seeding (CSG-UOO2002) in New Zealand.

\bmsection{Disclosures} The authors declare no conflicts of interest.

\bmsection{Data availability} Data underlying the results presented in this paper are available at Zenodo at Diamond-turned GaAs WGMR with million optical Q factor L.\ M.\ Platt, M.\ I.\ Suresh, F.\ Azeem, L.\ S.\ Trainor, and H.\ G.\ L.\ Schwefel, Zenodo 2024 (DOI: 10.5281/zenodo.8072306).


\begin{thebibliography}{10}
\newcommand{\enquote}[1]{``#1''}

\bibitem{haas_gallium_2020}
J.~Haas, R.~Stach, C.~Kolm, R.~Krska, and B.~Mizaikoff, \enquote{Gallium
  arsenide waveguides as a platform for direct mid-infrared vibrational
  spectroscopy,} {\protect\JournalTitle{Analytical and Bioanalytical
  Chemistry}} \textbf{412}, 3447--3456 (2020).

\bibitem{chen_on-chip_2020}
Y.~Chen, J.~Li, X.~Guo, L.~Wan, J.~Liu, Z.~Chen, J.~Pan, B.~Zhang, Z.~Li, and
  Y.~Qin, \enquote{On-chip high-sensitivity photonic temperature sensor based
  on a {GaAs} microresonator,} {\protect\JournalTitle{Optics Letters}}
  \textbf{45}, 5105--5108 (2020). Publisher: Optica Publishing Group.

\bibitem{geum_monolithic_2019}
D.-M. Geum, S.~Kim, S.~K. Kim, S.~Kang, J.~Kyhm, J.~Song, W.~J. Choi, and
  E.~Yoon, \enquote{Monolithic integration of visible {GaAs} and near-infrared
  {InGaAs} for multicolor photodetectors by using high-throughput epitaxial
  lift-off toward high-resolution imaging systems,}
  {\protect\JournalTitle{Scientific Reports}} \textbf{9}, 18661 (2019). Number:
  1 Publisher: Nature Publishing Group.

\bibitem{van_all-optical_2002}
V.~Van, T.~Ibrahim, K.~Ritter, P.~Absil, F.~Johnson, R.~Grover, J.~Goldhar, and
  P.-T. Ho, \enquote{All-optical nonlinear switching in {GaAs}-{AlGaAs}
  microring resonators,} {\protect\JournalTitle{IEEE Photonics Technology
  Letters}} \textbf{14}, 74--76 (2002). Conference Name: IEEE Photonics
  Technology Letters.

\bibitem{ravindran_gaas_2012}
S.~Ravindran, A.~Datta, K.~Alameh, and Y.~T. Lee, \enquote{{GaAs} based
  long-wavelength microring resonator optical switches utilising bias assisted
  carrier-injection induced refractive index change,}
  {\protect\JournalTitle{Optics Express}} \textbf{20}, 15610--15627 (2012).
  Publisher: Optica Publishing Group.

\bibitem{chin_gaas_1999}
M.~Chin, C.~Youtsey, W.~Zhao, T.~Pierson, Z.~Ren, S.~Wu, L.~Wang, Y.~Zhao, and
  S.~Ho, \enquote{{GaAs} microcavity channel-dropping filter based on a
  race-track resonator,} {\protect\JournalTitle{IEEE Photonics Technology
  Letters}} \textbf{11}, 1620--1622 (1999). Conference Name: IEEE Photonics
  Technology Letters.

\bibitem{bosi_potential_2007}
M.~Bosi and C.~Pelosi, \enquote{The potential of {III}-{V} semiconductors as
  terrestrial photovoltaic devices,} {\protect\JournalTitle{Progress in
  Photovoltaics: Research and Applications}} \textbf{15}, 51--68 (2007).
  \_eprint: https://onlinelibrary.wiley.com/doi/pdf/10.1002/pip.715.

\bibitem{bett_iii-v_1999}
A.~Bett, F.~Dimroth, G.~Stollwerck, and O.~Sulima, \enquote{{III}-{V} compounds
  for solar cell applications,} {\protect\JournalTitle{Applied Physics A}}
  \textbf{69}, 119--129 (1999).

\bibitem{wada_optoelectronic_1988}
O.~Wada, \enquote{Optoelectronic integration based on {GaAs} material,}
  {\protect\JournalTitle{Optical and Quantum Electronics}} \textbf{20},
  441--474 (1988).

\bibitem{wang_quasi-phase_2019}
S.~Wang, T.~Matsushita, T.~Matsumoto, S.~Yoshida, and T.~Kondo,
  \enquote{Quasi-phase matched difference frequency generation in
  corrugation-reduced {GaAs}/{AlGaAs} periodically-inverted waveguides,}
  {\protect\JournalTitle{Japanese Journal of Applied Physics}} \textbf{58},
  SBBE01 (2019). Publisher: IOP Publishing.

\bibitem{otman_phase_2018}
N.~A. Otman and M.~Cada, \enquote{Phase {Matching} for {Difference} {Frequency}
  {Generation} in {GaAs} {Via} an {Artificial} {Birefringence} {Technique}
  {Using} {Silver} {Nanowires},} {\protect\JournalTitle{IEEE Photonics
  Journal}} \textbf{10}, 1--10 (2018). Conference Name: IEEE Photonics Journal.

\bibitem{sotor_all-fiber_2018}
J.~Sotor, T.~Martynkien, P.~G. Schunemann, P.~Mergo, L.~Rutkowski, and
  G.~Soboń, \enquote{All-fiber mid-infrared source tunable from 6 to
  9 um based on difference frequency generation in {OP}-{GaP}
  crystal,} {\protect\JournalTitle{Optics Express}} \textbf{26}, 11756--11763
  (2018). Publisher: Optica Publishing Group.

\bibitem{fu_high-beam-quality_2019}
Q.~Fu, L.~Xu, S.~Liang, P.~C. Shardlow, D.~P. Shepherd, S.-U. Alam, and D.~J.
  Richardson, \enquote{High-beam-quality, watt-level, widely tunable,
  mid-infrared {OP}-{GaAs} optical parametric oscillator,}
  {\protect\JournalTitle{Opt. Lett., OL}} \textbf{44}, 2744--2747 (2019).
  Publisher: Optica Publishing Group.

\bibitem{becheker_optical_2022}
R.~Becheker, M.~Bailly, S.~Idlahcen, T.~Godin, B.~Gerard, H.~Delahaye,
  G.~Granger, S.~Fèvrier, A.~Grisard, E.~Lallier, and A.~Hideur,
  \enquote{Optical parametric generation in {OP}-{GaAs} waveguides pumped by a
  femtosecond fluoride fiber laser,} {\protect\JournalTitle{Optics Letters}}
  \textbf{47}, 886--889 (2022). Publisher: Optica Publishing Group.

\bibitem{alanis_optical_2019}
J.~A. Alanis, M.~Lysevych, T.~Burgess, D.~Saxena, S.~Mokkapati, S.~Skalsky,
  X.~Tang, P.~Mitchell, A.~S. Walton, H.~H. Tan, C.~Jagadish, and P.~Parkinson,
  \enquote{Optical {Study} of p-{Doping} in {GaAs} {Nanowires} for
  {Low}-{Threshold} and {High}-{Yield} {Lasing},} {\protect\JournalTitle{Nano
  Letters}} \textbf{19}, 362--368 (2019). Publisher: American Chemical Society.

\bibitem{zhang_ultralow_2021}
X.~Zhang, R.~Yi, N.~Gagrani, Z.~Li, F.~Zhang, X.~Gan, X.~Yao, X.~Yuan, N.~Wang,
  J.~Zhao, P.~Chen, W.~Lu, L.~Fu, H.~H. Tan, and C.~Jagadish, \enquote{Ultralow
  {Threshold}, {Single}-{Mode} {InGaAs}/{GaAs} {Multiquantum} {Disk} {Nanowire}
  {Lasers},} {\protect\JournalTitle{ACS Nano}} \textbf{15}, 9126--9133 (2021).
  Publisher: American Chemical Society.

\bibitem{grisard_quasi-phase-matched_2012}
A.~Grisard, E.~Lallier, and B.~Gérard, \enquote{Quasi-phase-matched gallium
  arsenide for versatile mid-infrared frequency conversion,}
  {\protect\JournalTitle{Opt. Mater. Express, OME}} \textbf{2}, 1020--1025
  (2012). Publisher: Optica Publishing Group.

\bibitem{gonzalez_second-harmonic_2013}
L.~P. Gonzalez, D.~C. Upchurch, P.~G. Schunemann, L.~Mohnkern, and S.~Guha,
  \enquote{Second-harmonic generation of a tunable continuous-wave
  {CO}$_{\textrm{2}}$laser in orientation-patterned {GaAs},}
  {\protect\JournalTitle{Opt. Lett., OL}} \textbf{38}, 320--322 (2013).
  Publisher: Optica Publishing Group.

\bibitem{tanimoto_quasi-phase-matching_2021}
R.~Tanimoto, Y.~Takahashi, and I.~Shoji, \enquote{Quasi-phase-matching stack of
  25 {GaAs} plates with high transmittance for high-power mid-infrared
  wavelength conversion fabricated by use of room-temperature bonding,}
  {\protect\JournalTitle{J. Opt. Soc. Am. B, JOSAB}} \textbf{38}, B30--B34
  (2021). Publisher: Optica Publishing Group.

\bibitem{kuo_4-quasi-phase-matched_2009}
P.~S. Kuo, W.~Fang, and G.~S. Solomon, \enquote{4-quasi-phase-matched
  interactions in {GaAs} microdisk cavities,} {\protect\JournalTitle{Optics
  Letters}} \textbf{34}, 3580--3582 (2009). Publisher: Optica Publishing Group.

\bibitem{kuo_second-harmonic_2014}
P.~S. Kuo, J.~Bravo-Abad, and G.~S. Solomon, \enquote{Second-harmonic
  generation using -quasi-phasematching in a {GaAs} whispering-gallery-mode
  microcavity,} {\protect\JournalTitle{Nature Communications}} \textbf{5}, 3109
  (2014). Number: 1 Publisher: Nature Publishing Group.

\bibitem{andronico_difference_2008}
A.~Andronico, I.~Favero, and G.~Leo, \enquote{Difference frequency generation
  in {GaAs} microdisks,} {\protect\JournalTitle{Optics Letters}} \textbf{33},
  2026--2028 (2008). Publisher: Optica Publishing Group.

\bibitem{chang_strong_2019}
L.~Chang, A.~Boes, P.~Pintus, J.~D. Peters, M.~Kennedy, X.-W. Guo, N.~Volet,
  S.-P. Yu, S.~B. Papp, and J.~E. Bowers, \enquote{Strong frequency conversion
  in heterogeneously integrated {GaAs} resonators,} {\protect\JournalTitle{APL
  Photonics}} \textbf{4}, 036103 (2019). Publisher: American Institute of
  Physics.

\bibitem{allain_electro-optomechanical_2021}
P.~E. Allain, B.~Guha, C.~Baker, D.~Parrain, A.~Lemaître, G.~Leo, and
  I.~Favero, \enquote{Electro-{Optomechanical} {Modulation} {Instability} in a
  {Semiconductor} {Resonator},} {\protect\JournalTitle{Physical Review
  Letters}} \textbf{126}, 243901 (2021).

\bibitem{kuo_on-and-off_2011}
P.~S. Kuo and G.~S. Solomon, \enquote{On- and off-resonance second-harmonic
  generation in {GaAs} microdisks,} {\protect\JournalTitle{Optics Express}}
  \textbf{19}, 16898--16918 (2011). Publisher: Optica Publishing Group.

\bibitem{kuo_mixing_2018}
P.~S. Kuo and M.~M. Fejer, \enquote{Mixing of polarization states in zincblende
  nonlinear optical crystals,} {\protect\JournalTitle{Optics Express}}
  \textbf{26}, 26971--26984 (2018). Publisher: Optica Publishing Group.

\bibitem{baker_critical_2011}
C.~Baker, C.~Belacel, A.~Andronico, P.~Senellart, A.~Lemaitre, E.~Galopin,
  S.~Ducci, G.~Leo, and I.~Favero, \enquote{Critical optical coupling between a
  {GaAs} disk and a nanowaveguide suspended on the chip,}
  {\protect\JournalTitle{Applied Physics Letters}} \textbf{99}, 151117 (2011).

\bibitem{yoon_gaas_2010}
J.~Yoon, S.~Jo, I.~S. Chun, I.~Jung, H.-S. Kim, M.~Meitl, E.~Menard, X.~Li,
  J.~J. Coleman, U.~Paik, and J.~A. Rogers, \enquote{{GaAs} photovoltaics and
  optoelectronics using releasable multilayer epitaxial assemblies,}
  {\protect\JournalTitle{Nature}} \textbf{465}, 329--333 (2010). Number: 7296
  Publisher: Nature Publishing Group.

\bibitem{wirths_room-temperature_2018}
S.~Wirths, B.~F. Mayer, H.~Schmid, M.~Sousa, J.~Gooth, H.~Riel, and K.~E.
  Moselund, \enquote{Room-{Temperature} {Lasing} from {Monolithically}
  {Integrated} {GaAs} {Microdisks} on {Silicon},} {\protect\JournalTitle{ACS
  Nano}} \textbf{12}, 2169--2175 (2018). Publisher: American Chemical Society.

\bibitem{brooks_integrated_2021}
A.~Brooks, X.-L. Chu, Z.~Liu, R.~Schott, A.~Ludwig, A.~D. Wieck, L.~Midolo,
  P.~Lodahl, and N.~Rotenberg, \enquote{Integrated
  {Whispering}-{Gallery}-{Mode} {Resonator} for {Solid}-{State} {Coherent}
  {Quantum} {Photonics},} {\protect\JournalTitle{Nano Lett.}} \textbf{21},
  8707--8714 (2021). Publisher: American Chemical Society.

\bibitem{guha_surface-enhanced_2017}
B.~Guha, F.~Marsault, F.~Cadiz, L.~Morgenroth, V.~Ulin, V.~Berkovitz,
  A.~Lemaître, C.~Gomez, A.~Amo, S.~Combrié, B.~Gérard, G.~Leo, and
  I.~Favero, \enquote{Surface-enhanced gallium arsenide photonic resonator with
  quality factor of 6E6,} {\protect\JournalTitle{Optica,
  OPTICA}} \textbf{4}, 218--221 (2017). Publisher: Optica Publishing Group.

\bibitem{floriansthesis}
F.~Sedlmeir, \enquote{Cystalline whispering gallery mode resonators,} Ph.D.
  thesis, University of Otago (2016).

\bibitem{chen_fundamental_2020}
J.~Chen, F.~Ding, X.~Luo, X.~Rao, and J.~Sun, \enquote{Fundamental study of
  ductile-regime diamond turning of single crystal gallium arsenide,}
  {\protect\JournalTitle{Precision Engineering}} \textbf{62}, 71--82 (2020).

\bibitem{azeem_active_2022}
F.~Azeem, \enquote{Active and passive crystalline whispering gallery mode
  resonators,} Thesis, University of Otago (2022). Accepted:
  2022-10-03T23:00:57Z.

\bibitem{couplingviaprism}
G.~A. Santamaría-Botello, L.~E.~G. Muñoz, F.~Sedlmeir, S.~Preu,
  D.~Segovia-Vargas, K.~A. Abdalmalak, S.~L. Romano, A.~G. Lampérez,
  S.~Malzer, G.~H. Döhler, H.~G.~L. Schwefel, and H.~B. Weber,
  \enquote{Maximization of the optical intra-cavity power of whispering-gallery
  mode resonators via coupling prism,} {\protect\JournalTitle{Optics Express}}
  \textbf{24}, 26503--26514 (2016).

\bibitem{near_fieldcoupling}
A.~Mazzei, S.~Götzinger, L.~S. Menezes, V.~Sandoghdar, and O.~Benson,
  \enquote{Optimization of prism coupling to high-q modes in a microsphere
  resonator using a near-field probe,} {\protect\JournalTitle{Optics
  Communications}} \textbf{250}, 428--433 (2005).

\bibitem{mccaulley_temperature_1994}
J.~A. {McCaulley}, V.~M. Donnelly, M.~Vernon, and I.~Taha, \enquote{Temperature
  dependence of the near-infrared refractive index of silicon, gallium
  arsenide, and indium phosphide,} {\protect\JournalTitle{Phys. Rev. B}}
  \textbf{49}, 7408--7417 (1994). Publisher: American Physical Society.

\bibitem{gorodetsky_rayleigh_2000}
M.~L. Gorodetsky, A.~D. Pryamikov, and V.~S. Ilchenko, \enquote{Rayleigh
  scattering in high-q microspheres,} {\protect\JournalTitle{J. Opt. Soc. Am. 
  B}} \textbf{17}, 1051--1057 (2000). Publisher:  Optica Publishing Group.

\bibitem{khan_optical_2007-1}
M.~J. Khan, J.~C. Chen, and S.~Kaushik, \enquote{Optical detection of terahertz
  radiation by using nonlinear parametric upconversion,}
  {\protect\JournalTitle{Optics Letters}} \textbf{32}, 3248 (2007).

\bibitem{lin_all-optical_2014}
Y.-C. Lin, M.-H. Mao, Y.-R. Lin, H.-H. Lin, C.-A. Lin, and L.~A. Wang,
  \enquote{All-optical switching in {GaAs} microdisk resonators by a
  femtosecond pump–probe technique through tapered-fiber coupling,}
  {\protect\JournalTitle{Optics Letters}} \textbf{39}, 4998--5001 (2014).
  Publisher: Optica Publishing Group.

\bibitem{roland_frequency_2020}
I.~Roland, A.~Borne, M.~Ravaro, R.~D. Oliveira, S.~Suffit, P.~Filloux,
  A.~Lemaître, I.~Favero, and G.~Leo, \enquote{Frequency doubling and
  parametric fluorescence in a four-port aluminum gallium arsenide photonic
  chip,} {\protect\JournalTitle{Optics Letters}} \textbf{45}, 2878--2881
  (2020). Publisher: Optica Publishing Group.

\bibitem{boikov_nonlinear_2024}
I.~K. Boikov, D.~Brunner, and A.~De~Rossi, \enquote{Nonlinear integrated
  optical resonators for optical fibre data recovery,}  (2024).
  ArXiv:2405.06102 [physics].

\end{thebibliography}

\end{backmatter}

\end{document}